\documentclass{evn2024}
\usepackage{graphicx,natbib}
\begin{document}
   \title{Jet properties of FR0 radio galaxies: need for VLBI data}

   \author{R. D. Baldi\inst{1}
          \and
          G. Giovannini\inst{1,2}
          \and
          A. Capetti\inst{3}
          \and
          R. Lico\inst{1,4}
          }

   \institute{INAF - Istituto di Radioastronomia, via Gobetti 101, 40129 Bologna, Italy
         \and
             Dipartimento di Fisica e Astronomia, Università di Bologna, Via P. Gobetti 93, I-40129 Bologna, Italy
             \and
             INAF - Osservatorio Astrofisico di Torino, Strada Osservatorio 20, I-10025 Pino Torinese, Italy
             \and
             Instituto de Astrof\'{\i}sica de Andaluc\'{\i}a-CSIC, Glorieta de la Astronom\'{\i}a s/n, 18008 Granada, Spain }

   \abstract{
Fanaroff-Riley (FR) type 0 radio galaxies are a subclass of radio-loud active galactic nuclei (AGN) that lack extended kpc-scale jets, different from the classical FRI and FRII radio galaxies. They constitute the most abundant population of radio galaxies in the local Universe (z$<$0.1), yet remain largely unexplored. VLBI observations of a limited number of FR0s demonstrated that their central supermassive black hole (SMBH) are able to lunch mostly two-sided jets, with mildly relativistic bulk speed. In this work, we highlight the need of further high-resolution radio observations to probe the jet structures of these compact radio galaxies, by showing exploratory results of our EVN+eMERLIN observation campaign of FR0s. A preliminary analysis of these recent data reveals a possible change of the jet direction at different scales. We shortly discuss their physical conditions to explain the observed jet compactness, stressing the role of the SMBH spin vector in shaping their radio morphology.
   }

   \maketitle
%

\section{Introduction}

Radio galaxies (RGs) are radio-loud active galactic nuclei (AGN) with misaligned relativistic jets that produce non-thermal emission observable across the electromagnetic spectrum. Typically associated with massive, gas-poor elliptical galaxies, RGs are classified into two categories: Fanaroff-Riley type I (FRI) and type II (FRII), based on radio morphology \citep{fanaroff74} and Low-Excitation or High-Excitation RGs based on accretion properties (low Eddington ratios $<$0.01 for LERGs, \citealt{best12,heckman14}). FRI sources exhibit edge-darkened jets and are generally LERGs, while FRII sources show edge-brightened jets and can be LERGs or HERGs \citep{baldi10b,miraghaei17,grandi21,mingo22}. Recently, large-scale surveys such as the 1.4-GHz FIRST \citep{first}, 150-MHz LOFAR \citep{shimwell22}, and SDSS optical surveys \citep{abazajian09} have revealed an enormous population of low-luminosity compact RGs ($<$10$^{24}$ W Hz$^{-1}$), termed FR0 galaxies, which lack extended jet emission \citep{baldi09,ghisellini11} and do not fit into the traditional FR I/II classification.

\begin{figure*}
   \centering
	\includegraphics[width=0.6\textwidth]{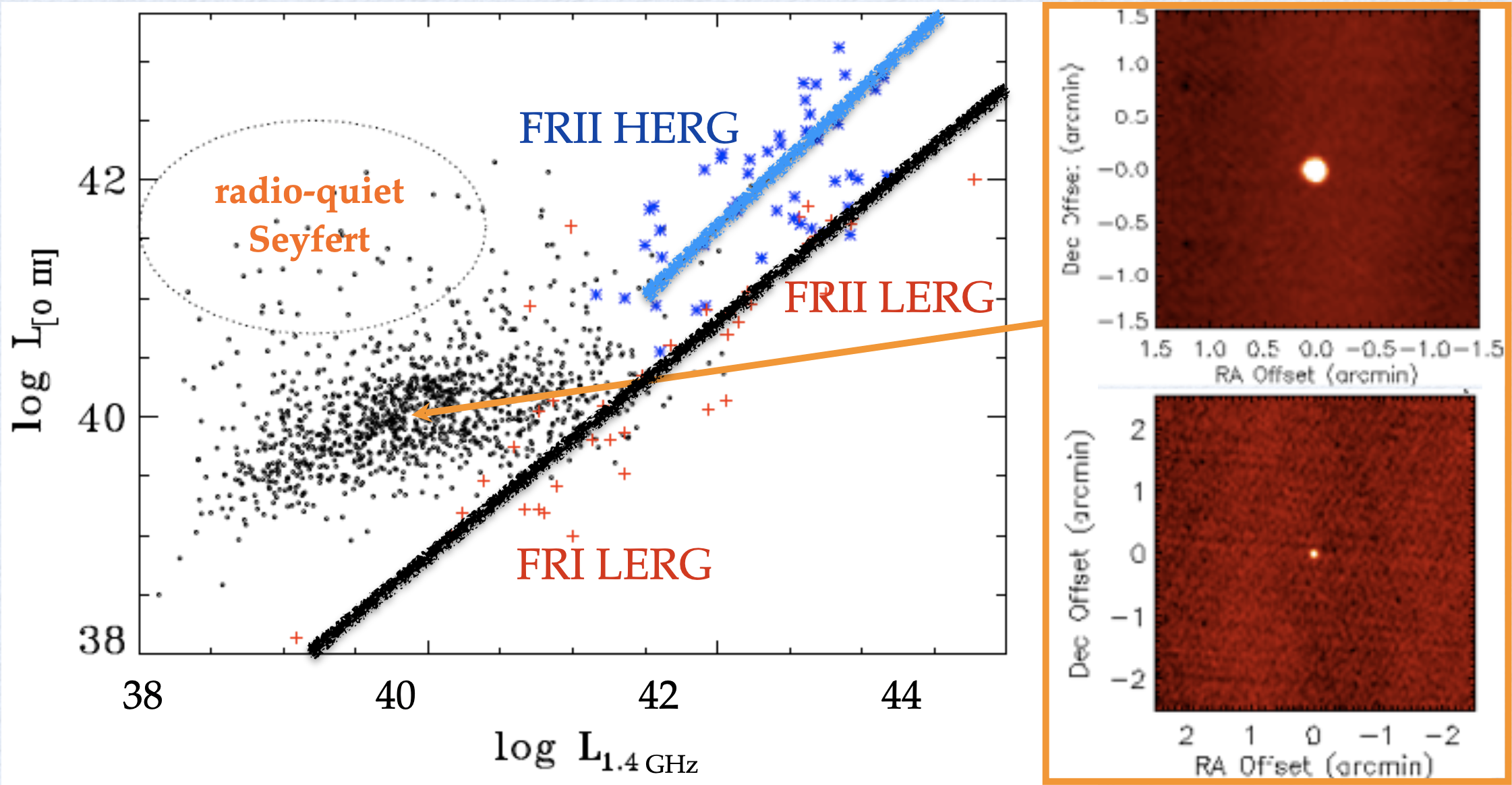}
        \caption[]{Accretion-ejection diagram: Total radio luminosity (1.4 GHz NVSS, \citealt{condon98}) vs. optical [O~III] line luminosity (SDSS, both in erg s$^{-1}$) as proxies for jet and accretion power in RGs at z $<$ 0.1. Dashed lines show the radio correlation for LERGs (black line, red pluses) and HERGs (blue line, blue asterisks) from the 3CR sample. The ellipse outlines the area of radio-quiet Seyferts \citealt{whittle85}. Right panels show typical compact radio structures in 1.4-GHz FIRST maps (5$\arcsec$ resolution), representing low-power FR0 RGs \citep{baldi23}.}
     
    \label{accretion-ejection-diagram}
\end{figure*}

In this context, \citet{best12} selected thousands of low-power RGs ($<$10$^{22}$ W Hz$^{-1}$) by cross-matching optical surveys, to search for early-type galaxies, and radio surveys (FIRST),  to find radio-band counterpart with flux densities $>$ 5 mJy, more than a factor of 1000 lower than the flux densities of classical FRI/FRIIs commonly considered for RG population studies (e.g., the Third Cambridge Catalog 3C \citealt{bennett62a,spinrad85}, $>$9 Jy at 178 MHz). One key tool to study the nature of this unexplored population is the accretion-ejection diagram (Fig.\ref{accretion-ejection-diagram}), where proxies for kinetic jet power  (L$_{\rm jet}$ from total radio power) and accretion disc power (L$_{\rm disc}$ from the [O~III] line luminosity) are used as axes (e.g. \citealt{heckman14} for a discussion). LERG and HERG populations mark two separate empirical correlations in this diagram, corresponding to two distinct disc-jet couplings, expressed by different ratios, L$_{\rm jet}$/L$_{\rm disc}$ $\sim$ -1.25 $-$ -0.5. The low-power RG population selected by \citet{best12} do not fully follow the LERG population unexpectedly, but show a larger scatter of jet power. More precisely, the low-power LERGs with extended jetted morphologies  ($\sim$20\%) in the FIRST maps follow the LERG correlation, while the majority show a deficit of total radio luminosities by a factor 100 $-$ 1000 with respect to the FRIs at the same accretion luminosity ($\sim$10$^{40}$ erg s$^{-1}$). This population appears to be all unresolved (compact) at a scale of 5 arcsec on the FIRST radio maps, which have been named FR0s \citep{baldi10b}. Their high core dominance is attributed to the lack of extended radio emission rather than an enhanced core, as they have similar core-to-emission line luminosity  and the similar nuclear properties compared to FRIs \citep{torresi18,baldi19,baldi23}.

Compact radio structure is characteristic of  several astrophysical sources. A bona-fide FR0 sample requires strict criteria: i) unresolved on kpc-scale, ii) red early-type galaxies, iii) large SMBH masses ($>$10$^{7.5}$ M$_{\odot}$), iv) LERG classification, minimizing spurious sources like star-forming and radio-quiet AGN. Accordingly, \citet{baldi18} selected 104 FR0s at $z<0.05$ with host-SMBH properties similar to LERGs \citep{heckman14}. A statistical study of the FR0 hosts  properties indicates that jet confinement or mechanical external frustration by the interstellar medium  is an unlikely scenario \citep{baldi23}.

\begin{figure*}
   \centering
	\includegraphics[width=0.7\textwidth]{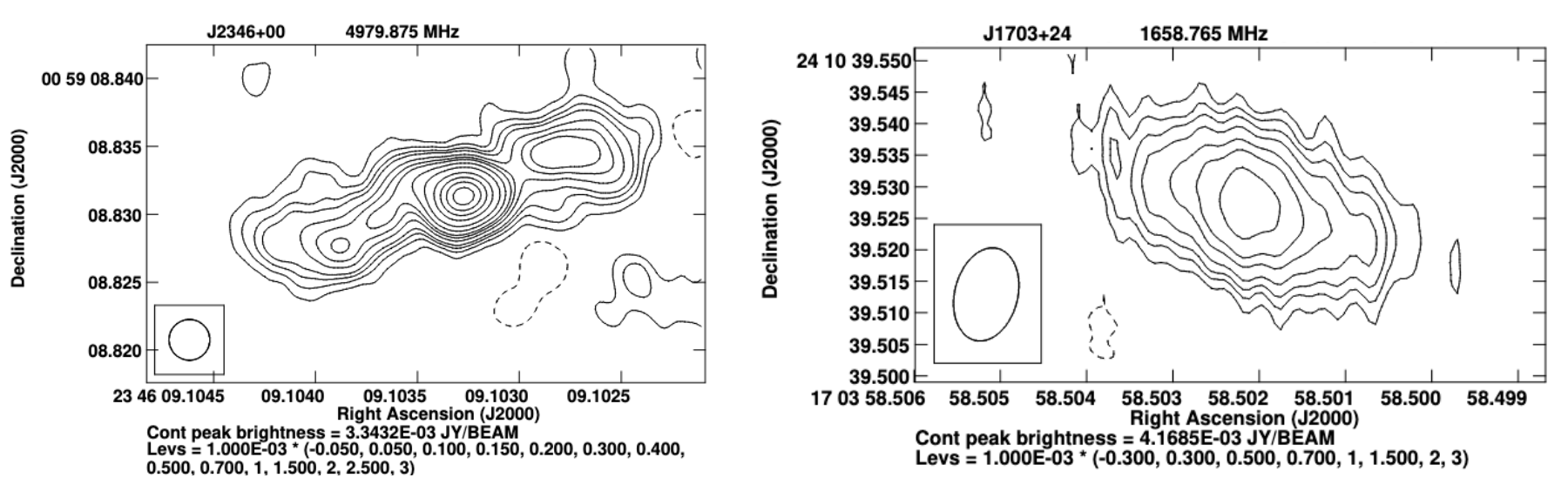}
        \caption[]{Examples of VLBI images of two FR0s in L and C band taken from \citet{giovannini23}.}
    \label{vlbifr0}
\end{figure*}

Compact, small radio structures are generally characteristic of young RGs, marking an early stage in their evolution \citep{snellen00,odea21}. A notable feature of young RGs (ages $<$10$^{6}$ yr) is their peaked radio spectra, caused by synchrotron self-absorption at low frequencies, without a flat spectrum at higher frequencies. This produces a peak at a specific turnover frequency, which decreases as the source matures \citep{odea97}. Conversely, FR0s show broader and less pronounced spectral curvatures from hundreds of MHz to GHz \citep{sadler14,capetti20} compared to genuinely young RGs like GHz-Peaked RGs. FR0s mostly exhibit flat or slightly convex spectra, with some showing inverted spectra at higher frequencies \citep{capetti19}. The high number density of FR0s (five times that of FR Is, \citealt{baldi23}) also challenges the interpretation of the entire FR0 class as young RGs, suggesting instead that additional considerations are needed to place them within the duty cycle and evolutionary scheme of the whole RG population. A large number of FR0s is also expected at high redshifts, given the significant identification of compact RGs observed in deep radio surveys (e.g., \citealt{bondi18,radcliffe21,vardoulaki21}).

\vspace{-0.5cm}
\subsection{Need for high-resolution radio observations}

 Multi-band Very Large Array (VLA) observations confirm that the majority of FR0 galaxies exhibit compact radio emission down to $\sim$0.3$\arcsec$, with only about 6 of the 25 sources resolved with kpc-scale jets \citep{baldi15,baldi19}. The enhanced Multi Element Remotely Linked Interferometer Network (eMERLIN) provides higher resolution down to 0.05$\arcsec$ Combining visibilities of VLA and eMERLIN (probing long and short baselines) reveals the presence of low-brightness sub-kpc jets in two cases, supporting the idea that extended  galactic-scale jets may be present in the whole FR0 population but are difficult to detect due to their faintness \citep{baldi21c}.

The VLBI (Very Long Baseline Interferometry) technique allows to access to pc-scale radio emission, offering crucial insights into FR0 jet properties near the jet-launching site. 
In fact, long-baseline FR0 observations turned out to be successful to detect low-brightness jet emission. Studies by \citet{cheng18} and \citet{cheng21}  using Very Long Baseline Array (VLBA) and  the European VLBI Network (EVN) resolved jets in 80\% of FR0s, indicating mildly relativistic jet speeds ($<$0.5c). However, these studies focused on particularly bright FR0s. \citet{giovannini23} have targeted less luminous FR0s: VLBA and EVN observations of 18 FR0CAT objects resoved pc-scale jets in most sources (Fig.~\ref{vlbifr0}).

High-resolution eMERLIN observations detected sub-mJy core components, with pc-scale emission contributing to 3–6\% of total kpc-scale emission.  Thirty targets have been observed so far with the VLBI technique (VLBA, EVN and eMERLIN). The majority are two-sided (15/30) sources and then one-sided (9/30) and a minority are unresolved (6/30) down to eMERLIN resolution of $\sim$50 mas. Furthermore, jet sidedness analysis suggests that the radio structures in FR0 galaxies are symmetric, confirming the mildly relativistic jet bulk speeds at pc scales, slower than classical FRIs \citep{giovannini23}.

 \begin{figure*}
   \centering
	\includegraphics[width=0.3\textwidth]{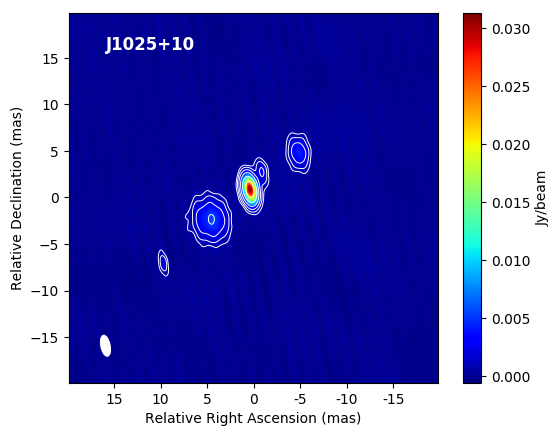}
 	\includegraphics[width=0.3\textwidth]{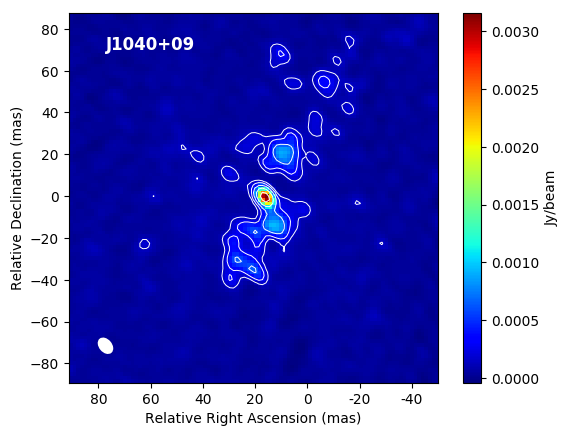}
	\includegraphics[width=0.3\textwidth]{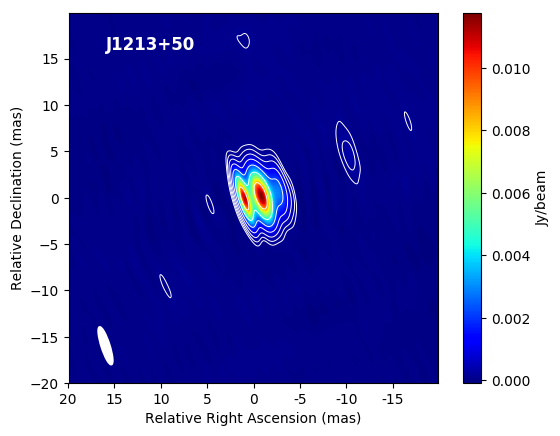}\\
 \includegraphics[width=0.3\textwidth]{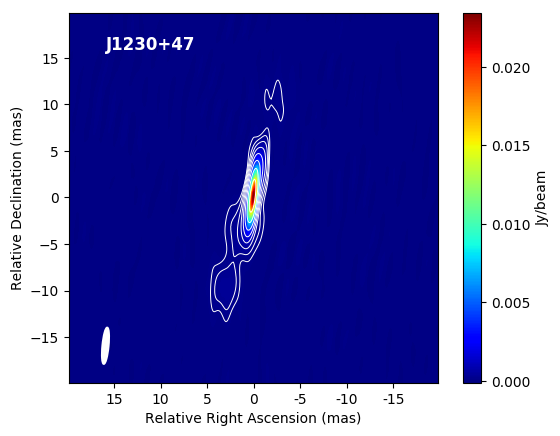}
 	\includegraphics[width=0.3\textwidth]{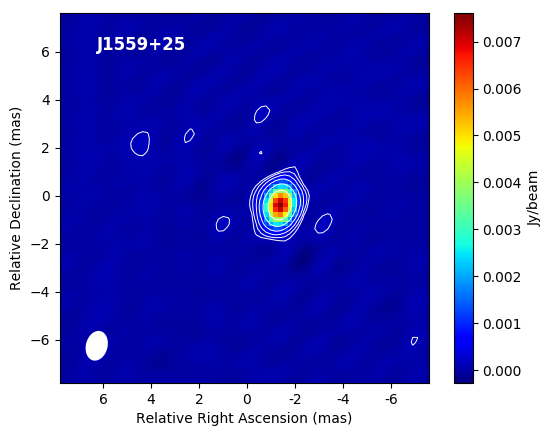}
        \caption[]{Preliminary calibrated maps of five (out of six) FR0s observed with EVN+eMERLIN array in C band (EG111 project). The contour levels are measured from the rms $\times$(1, 2, 4, 8, 16, 32, 64, 128), where the rms ranges between 30 and 80 $\mu$Jy. The beam is displayed as white ellipse in the left bottom corner.} 
    \label{evnemerlinfr0}
\end{figure*}

The current VLBI observations of FR0s demonstrated the power of this technique in resolving the jet structures of FR0s and highlighted the tremendous need for further high-resolution observations  to undertake a statistical study of the whole population.

\section{Data}

We have recently obtained EVN  observations (June 2022, including eMERLIN antenna, EG111 project) for a sample of 6 FR0s in C band in phase-reference mode. The EVN observations were performed with the following array (18 maximum and 9 minimum antennae available per session): Effelsberg, Jodrell Bank, Westerbork,  Medicina, Noto, Torun, Onsala, Hartebeesthoek, Irbene, Westerbork, Tianma65, Urumqi, Yebes, Pickmere, Knockin, Defford, Darnhall, Cambridge), with a maximum baseline of $\sim$130 k$\lambda$. A bandwidth of
64 MHz was used, divided into 8 intermediate frequency bands, all centred at 4.9 GHz. We obtained observation session of a total time of 4hr/source with at least 2hr on source. The data were processed at the EVN  correlator at the Joint Institute for VLBI in Europe (JIVE), in Dwingeloo,
The Netherlands. A first post-processing analysis was also carried out at JIVE. A first a priori visibility amplitude calibration was performed using antenna
gains and system temperatures measured at each antenna. A
fringe fitting  of the residual delays and fringe rates 
was performed for all the radio source using Effelsberg as reference antenna. The phase and amplitude calibration was performed with \verb|AIPS|. We then averaged the data in frequency and exported them to be imaged and self-calibrated in amplitude and phase in \verb|DIFMAP| with cell size of $\sim$0.3 mas for five sources. The final beam size has a large major axis of $\sim$1.3 mas (1 mas $\rightarrow$ 1 pc at z=0.05).

\vspace{-0.25cm}
\section{Preliminary results}

Figure ~\ref{evnemerlinfr0} depicts  the preliminary maps of five out of the six observed targets (J1025+10, J1040+09, J1213+50, J1230+47, J1559+25). All the sources have been detected with signal-to-noise  of higher than 5. The rms generally ranges between 30 and 80 $\mu$Jy beam$^{-1}$.

The morphologies are all two-sided structures, apart from J1559+25 which appears compact in full resolution (but resolved with VLBA by \citealt{cheng18}). The core flux densities range between 3 and 30 mJy beam$^{-1}$. The jetted emission appears extended on some tens of mas, which corresponds to some tens of parsec (Fig.~\ref{evnemerlinfr0}).

 \begin{figure*}
   \centering
		\includegraphics[width=0.7\textwidth]{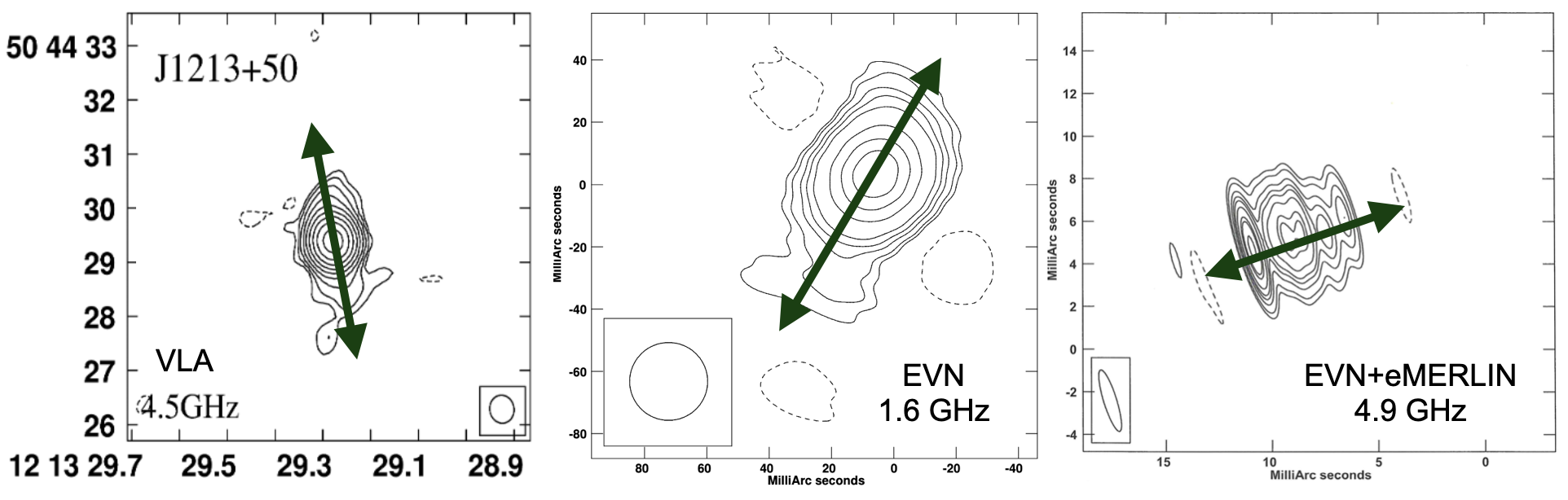}

        \caption[]{Panel of three sets of radio maps of the FR0 J1213+50 observed with VLA (4.5 GH, \citealt{baldi19}), EVN (1.6 GHz, \citealt{giovannini23}) and EVN+eMERLIN (new data, 4.9 GHz). The arrow roughly indicates the position angle of the jet direction, which changes a different physical scales across the maps.} 
    \label{jetchange}
\end{figure*}

A quick comparison between the radio maps obtained with different arrays (VLA, EVN, eMERLIN) reveals that in two targets the position angle of the extended emission, tracing the jet orientation, changes across the various maps. Figure~\ref{jetchange} show the case of J1213+50 whose radio emission elongation in the VLA map \citep{baldi19} on kpc scale is significantly different from that observed at the EVN scale (1.6 GHz, \citealt{giovannini23}) and new EVN+eMERLIN map observed on parsec scale. A more statistically robust analysis of this phenomenon will be performed in a forthcoming paper which will include the study of the whole EG111 program.

\vspace{-0.25cm}
\section{Conclusions}

Current VLBI studies on FR0 RGs provide an interesting insight on the jet properties of this abundant population of radio-loud AGN. A new EVN+eMERLIN high-resolution radio campaign confirms that FR0 {\it can} launch parsec-scale jets. The jet sidedness and symmetry and proper motion studies point to a mildly-relativist jet bulk speed, lower compared to FRIs and FRIIs. FR0 jets appear to be slower, with Lorentz factors 1$-$2 at parsec scale, suggesting that relativistic beaming effects are negligible in these sources. In addition, the evident re-orientation of the jet axis at different scale in two cases might indicate the possibility of a crucial role for the SMBH spin in keeping the radio source compact,  since jets are launched along the direction of the spin vectors \citep{stanghellini24}.

In fact, one possible explanation is that FR0 galaxies host SMBH with low prograde spin, which limits the energy available to launch powerful, extended jets based on a Blandford-Znajek stratified jet model. Over time, as accreting material increases the SMBH's angular momentum, some FR0s might evolve into FRI galaxies by increasing SMBH spin \citep{garofalo19}. Another possibility is that weak magnetic fields at the base of the jet could prevent the formation of large-scale structures, confining the jets to smaller sizes \citep{grandi21}. Numerical simulations of an FR0 with a conical jet relativistic profile show the development of a re-collimation  shock which promotes the growth of instabilities and a rapid jet deceleration, resulting in a compact low-power jet morphology \citep{costa24}.

Further VLBI studies are needed to investigate the jet launching/propagation mechanism in this class of intrinsically compact RGs. High-frequency (22-43 GHz) global VLBI observations would be ideal to resolve the inner jet structure. In this framework, the upgrade of the Italian VLBI antennae to the high-frequency band is, indeed, crucial for future studies. Furthermore, X-ray and optical-band studies, along with ad-hoc numerical simulations, will help to better quantify the impact of the FR0 jets on the hosts, and their relationship with the central SMBH properties and unearth the high-z population of FR0s at high redshifts.

\begin{acknowledgements}
\scriptsize

The European VLBI Network (www.evlbi.org) is a joint facility of independent European, African, Asian, and North American radio astronomy institutes. The scientific results from the data presented in this publication are derived from the following EVN project code: EG111.  R.D.B.
acknowledges financial support from INAF mini-grant \textit{\lq\lq FR0 radio galaxies\rq\rq} (Bando Ricerca Fondamentale INAF 2022).

\end{acknowledgements}

\scriptsize

\bibliographystyle{aa} 

\bibliography{my}

\end{document}